# WISE/NEOWISE observations of comet 103P/Hartley 2

James M. Bauer[1,2], Russell G. Walker[3], A. K. Mainzer[1], Joseph R. Masiero[1], Tommy Grav[4], John W. Dailey[2], Robert S. McMillan[5], Carey M. Lisse[6], Yan R. Fernández[7], Karen J. Meech[8], Jana Pittichova[8], Erin K. Blauvelt[1], Frank J. Masci[2], Michael F. A'Hearn[9], Roc M. Cutri[2], James V. Scotti[5], David J. Tholen[8], Emily DeBaun[1], Ashlee Wilkins[1], Emma Hand[1], Edward L. Wright[10], and the WISE Team

[1] Jet Propulsion Laboratory, California Institute of Technology, 4800 Oak Grove Drive, MS 183-401, Pasadena, CA 91109 (email: bauer@scn.jpl.nasa.gov)

[2] Infrared Processing and Analysis Center, California Institute of Technology, Pasadena, CA 91125

[3] Monterey Institute for Research in Astronomy, 200 Eighth Street, Marina, CA 93933

[4] Department of Physics and Astronomy, Johns Hopkins University, 3400 N. Charles Street, Baltimore, MD 21218

[5] Lunar and Planetary Laboratory, University of Arizona, 1629 East University Blvd., Kuiper Space Science Bldg. #92, Tucson, AZ 85721-0092

[6] Applied Physics Laboratory, Johns Hopkins University, 11100 Johns Hopkins Road Laurel, MD 20723-6099

[7] Department of Physics, University of Central Florida, 4000 Central Florida Blvd., P.S. Building, Orlando, FL 32816-2385

[8] Institute for Astronomy, University of Hawaii, 2680 Woodlawn Dr., Manoa, HI 96822

[9] Department of Astronomy, University of Maryland, College Park, MD 20742-2421

[10] Department of Physics and Astronomy, University of California, PO Box 91547, Los Angeles, CA 90095-1547

*Short Title: WISE IR observations - large grains & $CO_2$ emission.*

**Abstract**

We report results based on mid-infrared photometry of comet 103P/Hartley 2 taken during May 4-13, 2010 (when the comet was at a heliocentric distance of 2.3 AU, and an observer distance of 2.0 AU) by the Wide-field Infrared Survey Explorer (Wright et al. 2010). Photometry of the coma at 22 microns and data from the University of Hawaii 2.2-m telescope obtained on May 22, 2010 provide constraints on the dust particle size distribution, dlogn/dlogm, yielding power-law slope values of alpha = -0.97 +/- 0.10, steeper than that found for the inbound particle fluence during the Stardust encounter of comet 81P/Wild 2 (Green et al. 2004). The extracted nucleus signal at 12 microns is consistent with a body of average spherical radius of 0.6 +/- 0.2 km (one standard deviation), assuming a beaming parameter of 1.2. The 4.6 micron-band signal in excess of dust and nucleus reflected and thermal contributions may be attributed to carbon monoxide or carbon dioxide emission lines and provides limits and estimates of species production. Derived carbon dioxide coma production rates are $3.5(+/- 0.9) \times 10^{24}$ molecules per second. Analyses of the trail signal present in the stacked image with an effective exposure time of 158.4 seconds yields optical-depth values near $9 \times 10^{-10}$ at a delta mean anomaly of 0.2 deg trailing the comet nucleus, in both 12 and 22 μm bands. A minimum chi-squared analysis of the dust trail position yields a beta-parameter value of $1.0 \times 10^{-4}$, consistent with a derived mean trail-grain diameter of 1.1/ρ cm for grains of ρ g/cm$^3$ density. This leads to a total detected trail mass of at least $4 \times 10^{10}$ ρ kg.

**Introduction**

In our solar system, only five cometary bodies have been studied up close via spacecraft encounters. A careful characterization of these few bodies and their behavior over time therefore provides vital ground-truth tests to the interpretations drawn for the larger data set of cometary bodies obtained from telescopic observations. The EPOXI/DIXI encounter of 103P/Hartley 2 (103P) provided a unique view of a Jupiter-family comet (JFC) near its perihelion, when it was most active. The extended ground-based monitoring of 103P (Meech et al. 2011) indicated that activity increased over the course of its much-studied perihelion approach. Here we present a unique collection of combined data during its more sparsely sampled pre-perihelion period in the spring of 2010, over 5 months prior to the EPOXI encounter, when the comet was at nearly twice the distance from the Sun and was imaged by the Wide-field Infrared Survey Explorer mission (WISE; Wright et al. 2010).

The WISE mission surveyed the sky at four mid-IR wavelengths simultaneously, 3.35 μm (W1), 4.60 μm (W2), 11.56 μm (W3) and 22.08 μm (W4), with approximately one hundred times improved sensitivity over the IRAS mission (Wright et al. 2010). As part of an enhancement to the WISE data processing system called "NEOWISE", funded by NASA's Planetary Science Division, the WISE Moving Object Processing Software (WMOPS) was developed to find solar system bodies in the WISE images. WMOPS successfully found a wide array of primitive bodies, including Near-Earth Objects (NEOs), Main Belt asteroids, comets, Trojans, and Centaurs. As of January 2011, NEOWISE has identified more than 157,000 small bodies, including 123 comets (Mainzer et al. 2011A). Such infrared observations are useful for determining size and albedo distributions, thermo-physical related properties such as thermal inertia, the magnitude of non-gravitational forces, and surface roughness (Mainzer et al. 2011B). The subset of these bodies exhibiting cometary activity require special treatment in the interpretation of such observations, owing to the material surrounding the solid nucleus, i.e. the contribution to the IR flux from the gas and dust, and to the variable nature of the observed brightness of the object attributable to outbursts. These IR imaging data provide unique opportunities to characterize four main components of comets: the coma dust and gas, the nucleus, and the extended dust trail. The subject of this paper, 103P, was of particular interest owing to the EPOXI spacecraft flyby scheduled in November 2010. As a result, much supporting data, including visual-band wavelength observations near to the time of the WISE observations, and a well-characterized light curve at similar distances, enhance the value of the WISE data. The mid-IR data in turn augment the more comprehensive picture of the EPOXI/DIXI mission target by providing a detailed view of its behavior at an earlier epoch. The WISE observations occurred at a heliocentric distance over twice that at the time of the November 2010 flyby. 103P also proved an excellent test case for the array of information the WISE data set has provided for the total set of comets.

## Observations

In anticipation of the observation of multiple targets of interest, we applied for telescope time at several sites to obtain visual wavelength imaging and photometry near to the times of the WISE observations. A combined campaign of ground-based and space-based telescopes have provided characterization of 103P from the previous perihelion passage through the time of the encounter, and beyond (cf. Meech et al. 2011). Our team concentrated our telescope allocation time requests around the expected dates for the observation of 103P, but continued to observe 103P beyond that period as well. Table 1 lists the times of observations at the various ground based sites, in addition to the WISE observations obtained on May 10-11, 2010 (see Figure 1). A total of 8 nights of observation are listed, characterizing the behavior of the comet both at the time of the WISE observations and near the comet's perihelion and encounter times (October 20 and November 4, 2010, UT, respectively).

Table 1: Mid-IR and ground-based observations of 103P

| Date/Time(UT) | Telescope/Instrument[1] | Wavelength (μm) | Images | Exposures[2] | Comments |
|---|---|---|---|---|---|
| 17Apr2010, 12:02 | Pal. 200", LFC | 0.7 | 3 | 60 | Poor Seeing, Thin Cirrus |
| 18Apr2010, 10:03 | SOAR 4.1m, SOI | 0.7 | 3 | 60 | Photometric |
| 23May2010, 13:52 | UH 2.2m, Tek2048 | 0.7 | 8 | 300 | Photometric |
| 13Jul2010, 09:30 | Pal. 200", LFC | 0.7 | 4 | 30 | Photometric |
| 17Aug2010, 03:56 | SOAR 4.1m, SOI | 0.7 | 5 | 180 | Photometric |
| 03Sep2010, 06:46 | Pal. 200", LFC | 0.7 | 3 | 60 | Photometric |
| 07Nov2010, 09:35 | Pal. 200", LFC | 0.7 | 22 | 20 | Photometric |
| 27Jan2011, 05:58 | Steward 2.3m, 90prime | 0.7 | 1 | 15 | Thin Cirrus |
| 10May2010, 08:18 – 11May2010, 11:59[3] | WISE | 3.4,4.6,11.6, 22.1 | 18 | 8.8 | Coma & Nucleus |
| 04May2010, 06:13 – 13May2010, 04:33 | WISE | 11.6, 22.1 | 113 | 8.8 | Debris Trail observations |

[1] Pal 200" = The Palomar 200-inch telescope, Mt. Palomar, CA., Southern Astronomical Research Telescope, Cerro Pachon, Chile, UH 2.2m = The University of Hawaii 2.2 meter telescope, Steward 2.3m = the Steward Observatory 2.3 meter telescope on Kitt Peak, with the 90primeOne camera. [2] Average Exposure Time in seconds. [3] individual WISE exposure Modified Julian Date values: 55326.34566, 55326.47797, 55326.61027, 55326.74258, 55326.87488, 55326.94097, 55327.00718, 55327.07327, 55327.13949, 55327.20558, 55327.27179, 55327.33788, 55327.40410, 55327.47018, 55327.60249, 55327.73479, 55327.86710, 55327.99940.

A total of 18 four-color WISE images (Figure 1) were obtained of the nucleus and coma over the course of 39.7 hours. That this was more than the average 10 WISE images obtained per sky object was attributable to the largely pro-grade motion of the comet (45 to 70 arcseconds/hr) moving in the direction of the scan progression of the spacecraft (WISE covers all ecliptic latitudes each day in a narrow band of sky at 90 +/- 2 deg elongation, and uses the spacecraft orbital motion around the Sun to scan this band across all ecliptic latitudes over 1 year). Variation in the sky-motion rates were attributable mostly to the spacecrafts orbital motion; in any case this reflex motion created a maximum blurring of ~0.02 arcseconds, an insignificant factor in the imaging, as the blur was <1% of the PSF FWHM in the shortest wavelength (2.75 arcseconds/pixel average scale of the images and 5.5 arcseconds/pixel in the 2×2-binned 22μm images ; Wright et al. 2010). The long, extended debris trail was imaged piecewise over the course of 9 days, using a total of 113 WISE images.

**Analysis**

*Optical Photometry:* For all but the night of April 17, 2010 and January 27, 2011 the Palomar, SOAR, Mauna Kea and Steward observatory ground-based observing conditions for 103P were photometric, and provided magnitudes to within a few percent uncertainty. After acquisition, the ground-based images were bias-subtracted and flat-field corrected. The R-band photometry obtained from the ground-based observations is summarized in Table 2, including flux magnitudes and geometry corrected A$f\rho$ dust column values (A'Hearn et al. 1984) as well as magnitudes. The fixed 2.5 and 5 arcsecond aperture radius values listed in Table 2 are chosen to highlight the activity near the comet's coma center. The values represent the average of the aperture signals in the individual images, but special analysis was performed for the April 18 and May 23, 2010 (Figure 2) observations. Larger aperture radius values of 11 and 22 arcseconds were used to match the aperture sizes necessary to obtain the signal from the poorest resolution WISE bands. We stacked and median-combined the May 23$^{rd}$ UH frames into a single image to reduce the contribution of background stars to the large aperture and background annulus signals. Photometric analysis yielded a reflected-light signal in these large aperture values for comparison to the WISE observations by adjusting the April 18 and May 23, 2010 aperture magnitudes to the May 10$^{th}$ WISE spacecraft-to-comet and comet heliocentric distance. Interpolation between the adjusted 18 April and 23 May, 2010 magnitudes yielded May 10, 2010 magnitudes of 18.45 +/- 0.02, and 18.11 +/- 0.02, and Afρ values of 0.85 and 0.69, for the 11 and 22 arcsecond aperture signal, respectively. Note that the interpolation correction was 0.04 mag, within the photometric uncertainty of the April 18 magnitudes and on the order of what might be expected for the change in viewing phase from 26.1 to 27.3 degrees for a cometary body (with slope parameters between 0.02 and 0.035 mag/degree; Meech & Jewitt 1987). A single V-band 300-second exposure was taken the night of the May 23, 2010 observations, yielding V-R colors of 0.38 +/- 0.03, or nearly solar.

Table 2: R-band Optical Photometry

| Date/Time (UT) | Airmass | $m_R$, $r_{ap}$=2.5" | $m_R$, $r_{ap}$=5" | Afρ, 2.5" [log cm] | Afρ, 5" [log cm] | $R_{helio}$ / $\Delta$ |
|---|---|---|---|---|---|---|
| 17Apr, 12:02 | 1.70 | 19.90 +/- 0.12 | 19.49 +/- 0.19 | 1.08 | 0.94 | 2.49 / 2.52 |
| 18Apr, 10:03 | 1.20 | 20.03 +/- 0.04 | 19.62 +/- 0.06 | 1.02 | 0.89 | 2.47 / 2.49 |
| 23May, 13:52 | 1.12 | 19.46 +/- 0.01 | 19.15 +/- 0.01 | 1.00 | 0.82 | 2.19 / 1.81 |
| 13Jul, 09:30 | 1.08 | 17.45 +/- 0.02 | 16.87 +/- 0.02 | 1.32 | 0.98 | 1.73 / 0.94 |
| 17Aug, 03:56 | 2.46 | 15.99 +/- 0.01 | 15.27 +/- 0.01 | 1.49 | 1.05 | 1.43 / 0.52 |
| 03Sep, 06:46 | 1.03 | 15.26 +/- 0.02 | 14.54 +/- 0.01 | 1.54 | 0.99 | 1.29 / 0.38 |
| 07Nov, 09:35 | 1.44 | 13.60 +/- 0.03 | 12.55 +/- 0.01 | 1.70 | 1.02 | 1.07 / 0.17 |
| 27Jan, 05:58 | 1.34 | 15.28 +/- 0.20 | 14.63 +/- 0.20 | 1.98 | 1.94 | 1.59 / 0.69 |

*WISE Observations:* The coma images were identified using the method described in the WISE archive explanatory supplement (http://wise2.ipac.caltech.edu/docs/release/prelim/expsup/), utilizing the WMOPS detections in combination with the single-frame (level 1b) archive source search. An additional frame with a near-edge detection (Modified Julian Date 55327.47018; see Table 1) was found from individually searching the archive scans spanning the dates of coverage (May 10-11, 2010).

The WISE data were processed through the WISE Science Data System (WSDS) pipeline (Wright et al. 2010), to remove detector artifacts as well as bias and flat correct each image. The resultant images were stacked into the comet-centered images shown in Figure 1, using the WSDS co-adder routine specially adapted to stack according to the comet's rate of apparent sky motion. Aperture photometry was performed on the stacked images of each band for apertures using radii of 6, 9, 11 and 22 arcseconds. Counts were converted to fluxes using the band-appropriate magnitude zero-points and $0^{th}$ magnitude flux values provided in Wright et al. (2010), and an iterative fitting to a black-body curve was conducted on the two long-wavelength bands to determine the appropriate color correction as listed in the same. Figure 3 shows the black body fit and the signal in W2, W3 and W4. The W1 image did not yield significant signal in any of the apertures. However, the reflected light signal was constrained by the visual wavelength data also shown on the graph. In the Figure 3 fluxes, the flux contribution from the nucleus (see Figure 4) has been removed. Even so, note that the flux in W2 is significantly greater than the total dust reflected light and thermal contribution, suggesting another narrow-band source of emission in the 4 – 5 um range of the W2 passband.

Table 3: Coma Mid-IR Fluxes (mJy)

| $R_{Aperture}$ | 4.6μm | 12μm | 22μm |
|---|---|---|---|
| 6" | 0.16 +/- 0.02 | 2.2 +/- 0.3 | 5 +/- 1 |
| 9" | 0.29 +/- 0.03 | 4.2 +/- 0.6 | 11 +/- 2 |
| 11" | 0.39 +/- 0.04 | 5.4 +/- 0.7 | 15 +/- 2 |
| 22" | 0.86 +/- 0.05 | 10 +/- 1 | 31 +/- 5 |

In order to extract the nucleus signal, we used routines developed by our team (Lisse et al. 1999 & Fernandez et al. 2000) to fit the coma as a function of angular distance from the central brightness peak along separate azimuths. The extracted nucleus signal in W3 and W4, 0.7 +/- 0.2 mJy and 2.6 +/- 0.7 mJy respectively, were fit to a NEATM model (Harris 1989, Delbo et al. 2003, & Mainzer et al. 2011B) with free and fixed beaming (η) parameters (Figure 4). Fits with η values fixed to 2.0, 1.2 (Stansberry et al. 2007) and 0.94 (Fernandez et al. 2008) yielded diameters of 1.5, 1.1 and 0.9 (+/- 0.2) km. The geometric mean radius, based on radar measurements (Harmon et al. 2010), is closer to 0.58 km, or a diameter of 1.16, more consistent with the η=1.2 fixed value, and are closer to the 1.14 km diameter value derived from Spitzer measurements of the 13 Aug 2008, $r_h$ = 5.4 AU, 22 um flux obtained by Lisse et al. (2009). The η=1.2 fixed fit yields an albedo of $p_v$= 0.038 +/- 0.004, assuming an absolute magnitude of 18.8 consistent with a 0.6 km mean radius (Lisse et al. 2009). Note that each fit iteration requires an interpolation for surface temperature in the WISE bands (Wright et al. 2010), so that different flux values are derived for each final fit, as shown on Figure 4. The interpolated corrections for temperature are largest in W3. For decreasing η, the temperature increases. Note that the flux contribution from the nucleus in W2 was calculated to be less than $10^{-5}$ Jy, or significantly less than the uncertainty in the 4.6μm flux. The W3 and W4 flux contributions from the nucleus were also subtracted from the total fluxes used to calculate the particle size distribution shown in Figure 5.

*Dust Trail:* Observations of the 103P/Hartley 2 debris trail spanned the time period from May 4:13:35 to May 13:33:47, and a position on the sky of $1.3^o$ leading to $4^o$ trailing the comet, covering a range from $-0.32^o$ to $+1.08^o$ in delta Mean Anomaly (dMA). The debris trail of 103P is a low contrast feature with respect to the infrared background, and requires co-addition to obtain reasonable signal-to-noise. The co-addition should be of the same trail features, for example, at the same dMA, to preserve any variations in the emission history.

W3 and W4 images were selected from the set of 113 WISE level 1b frames containing the trail, and the W3 images re-binned to the W4 pixel size. The background radiance of each image was fit with a two dimensional low order polynomial and subtracted from the image to reduce the effects of global radiance slopes. The resulting images were then shifted to bring the desired dMA to the center of a stacked (co-added) image of 508 columns x 200 rows, and rotated about that dMA to bring the projected orbit of 103P parallel to the stacked image rows. The co-added image thus contains the orbit along row 100 with the selected dMA at the center. The increment of dMA used was $0.04^o$, and the number of images stacked at each dMA varied from 14 to 19. Figure 6A shows co-added images for a stack of W3 and W4 frames centered at dMA=$0.2^o$.

Radiance profiles normal to the trail were constructed from the mean and standard deviation of each row. Outlier rejection (of stars, instrumental artifacts, cosmic rays, etc.) was applied to each row using an iterative technique based on the theory of order statistics (Mandell,1964). The radiance profiles were fit to a Moffat (1969) function with an exponent of 0.5 to find the peak radiance, the position of the peak, and the full width at half maximum. Radiance profiles derived from the two images shown in Figure 6A are plotted in Figure 6B. The Moffat function, which is a modified Lorentz function, was chosen since it fits the peak, position, and shape of the radiance profile more accurately than the conventional Gaussian.  The Lorentz form of the trail profile was originally suggested as a result of dynamical modeling of trail grains (Lien, 1990). DN were converted to fluxes using the band-appropriate magnitude zero-points and zero magnitude flux values provided in Wright et al. (2010).  The peak profile in-band radiances are plotted as a function of dMA in Figure 7.  In this and subsequent figures, vertical dotted lines delineate a region in which the trail radiances are influenced by the coma and tail radiances. This is due, in part, to the small angle between the sky-projected anti-solar vector and the projected anti-velocity vector (see Figure 1). However, it is mainly a result of the "boxcar-like"  averaging effect of taking row means. It is also responsible for the peak radiance occurring at dMA = 0.16 rather than at the nucleus (dMA = 0) since the trail is brighter behind than in front of the comet. The mean radiance of the trail leading the comet is 0.05 MJy/sr in W4 and 0.02 MJy/sr in W3. Behind the comet, the trail peak radiance falls from 0.14 MJy/sr at 0.2° dMA to 0.06 MJy/sr at 1.04° dMA in W4 and 0.06 MJy/sr to 0.04 MJy/sr in W3.

**Discussion**

The three-color composite image and individual band images in Figure 1 show many of the summary features of the data set that provide unique characterization of 103P. The composite image shows an extended trail with thermal emission long-ward of 10 μm and with little or no reflected-light component in the shorter wavelength 3.4 and 4.6 μm bands. The trail also seems to be noticeably missing from the deep-exposure, stacked UH 2.2m image in Figure 2 taken within 12 days of the WISE data. There is also the hint of a dust trail in the 11.6 and 22.1 μm bands that precedes the comet, indicative of large grains remaining from the comet's previous passage. Note in the images that Hartley 2's trail-ward brightness variations do not persist in the stacked subsets of the images, and are likely caused by noise fluctuations rather than large-scale ejecta. At 4.6 μm, the image is considerably compact, even when compared to the R-band image from the UH 2.2m, suggesting the dominant source of the brightness is not the same source as the reflected light or thermal emission of the dust, as is likely in the WISE bands W3 & W4, and the R-band.

The aperture photometry at 11.6 and 22.1 μm, in combination with the R-band photometry, provides constraints on the dust particle size distribution (PSD). Proceeding as in Bauer et al. (2008), if we assume, starting with the longest wavelengths, that the thermal emission comes primarily from the dust particles near the size of the wavelength of emission, we can obtain an estimate of the number of large particles that fall within the aperture by scaling to the nucleus signal using the formula:

$$n_g = (R / a)^2 \times (F_{total, \lambda} - F_{nucl, \lambda}) / F_{nucl, \lambda} \qquad (1)$$

where $n_g$ is the total count of grains of radius $a$ (usually equal to ½ the observing wavelength $\lambda$), and $F_{total,\lambda}$ is the total flux, including the nucleus, while $F_{nucl,\lambda}$ is the flux contribution from the nucleus. For the shorter thermal wavelengths, the same calculation was repeated, but the contribution to the flux from the larger particles, as estimated from the derived number of particles and projected area for the assumed particle sizes, is subtracted off. The reflected light contribution from the larger particles is also subtracted from the short-wavelength flux before deriving a similar particle number from:

$$n_g = ((Af\rho \times \rho/p) / a)^2 \qquad (2)$$

where $Af\rho$ and $\rho$ are as defined in A'Hearn et al. (1984), p is the estimated particle reflectance (here 0.03), and $a$ is as in eq. (1). The number of particles as a function of particle mass is shown in the Figure 5 log-log plot, while derived values are shown in Table 3. Production rates for the dust sizes were calculated from the crossing time of the projected distance of the aperture (i.e. $\rho$, which is 15994 and 31889 km for 11 and 22 arcseconds, respectively), assuming an ejection velocity ~ 1km/s. Note that the dust production rate,

$$Q_{dust} = n_g \times m_g \times \rho/v_g \qquad (3)$$

is inversely proportional to the grain velocity, $v_g$, which is estimated here as similar to the $CO_2$ gas ejection velocity at this distance (cf. Pittichova et al. 2008). The number of coma dust grains, the dust mass estimates, and production rates as a function of size are summarized in Table 4. Log-$m_g$ vs. Log-$n_g$ is shown in Figure 5, and provides approximate particle size distribution power-law slopes ($\alpha$) for the dust particle size distribution (dust PSD). For comparison, dust PSD data are shown from various measured cometary bodies, including the Deep Impact ejecta from the impactor experiment on 9P/Tempel 1, the active Centaur Echeclus, and the Stardust Mission target 81P/Wild 2. Our estimate of dust PSD for 103P yields a moderately steeper slope ($\alpha = -0.97 +/- 0.08$) when compared to these other bodies, suggesting, when compared to the smaller grains, relatively fewer large-grained dust particles were generated while the comet was at a distance of 2.3 AU. Furthermore, the 11.6 and 22.1 μm data points suggest the 5.7 μm radius-scale grains are fewer than predicted by the simple power law. This may be indicative of a bifurcation in the distribution, where the mass is more strongly weighted to grain sizes greater than radii 5.7 μm, and may be consistent with the strong trail signal seen for 103P. Note that alternatively in the literature $\alpha$ is expressed as a size-dependant slope, rather than mass-dependant (cf. Fulle et al. 2004). The value quoted above translates into a $\alpha_{size} \approx 3.0 +/ 0.3$ (log-N/log-$a$), not an uncommon value for JFC comets that exhibit strong dust trails and weak silicate features, both indicative of abundant large dust grains (Lisse et al. 1998, Fulle et al. 2004). The total mass loss in coma dust emission over the course of an orbit at the observed rate would be $\sim 2 \times 10^9$ kg, assuming 1 g/cm$^3$ grain densities, or a rate of ~7 kg/s, and is consistent with the values derived by Lisse et al. (2009).

Figure 3 shows the thermal and reflected light contributions from the dust to the measured flux, after the black body temperature fit. The nucleus signal, insignificant compared to the flux uncertainty, has also been removed from each band. A neutral-reflectance for the coma grains was used originally in the computation for the combined thermal and reflected light dust contribution, owing to the near-solar V-R coma colors found using the UH data set. However, 103P was reported to have a noticeably red coma at the time of the encounter (cf. Sitko et al. 2011), into the near-infrared, prompting us to show, for comparison, a coma reflected-light model with a reddening law based on Jewitt & Meech (1986) averaged out to 3.5 µm, i.e. with dust reflectivity proportional to the wavelength to the approximately 0.2 power. The difference in the flux in W2 between the neutral-reflectance and reddening law reflected-light scaling models is well within the reported photometric uncertainty. However, we chose to base further calculations on the original neutral-reflectance model based on our UH measurements.

*4 um Anomalous Emission:* Comet 103P's signal at 4.6µm is about 0.4 mJy for an 11 arcsecond aperture. The contribution from the thermal and reflected light of the coma was calculated to be 0.03 mJy (see Figure 3). A large excess flux in the 4.6 µm band suggests other contributions, possibly from emission lines. Recent work with the Spitzer Space Telescope IRAC instrument's similar band-pass centered at 4.8 µm shows the most likely candidate emission lines to be from CO and $CO_2$ lines (Pittichova et al. 2008, Reach et al. 2009). With a band-pass of ~1µm (Wright et al. 2010), the 4.6 µm band contains both the CO (4.26 µm) and $CO_2$ (4.67µm) mid-IR emission bands. Water emission lines appropriate for the measured temperatures ~200K that reside in the mid-IR nearest the band pass cluster in the 2.7 – 33 and 5.5 – 6.3 um range (Lisse et al. 2006, Woodward et al. 2007). The excess of the W2 flux is likely attributable to CO and $CO_2$ gas species, seen strongly in Hartley 2 by the EPOXI mission in 2010 (A'Hearn et al. 2011) and in previous apparitions (Weaver et al. 1994; Crovisier et al. 1997), while CO was found to be very under-abundant in this comet and also in other comets as the dominant lines in the 4-5µm spectral range by ISO (Crovisier et al. 1999, Woodward et al. 2008, Reach et al. 2009). Comet 103P's signal at 4.6µm is about 0.4 mJy for an 11 arcsecond aperture. The contribution from the thermal and reflected light of the coma was calculated to be 0.03 mJy (see Figure 3).

Table 4: Production Rates

| Quantity | 12µm, 11" | 12µm, 22" | 22µm, 11" | 22µm, 22" | 4.6µm 11"/22" | | R-band, 11" | R-band, 22" |
|---|---|---|---|---|---|---|---|---|
| $m_g$[kg] | $8.1 \times 10^{-13}$ | | $5.7 \times 10^{-12}$ | | -- | -- | $1.8 \times 10^{-16}$ | |
| $n_g$ | $2 \times 10^{16}$ | $3 \times 10^{16}$ | $1.26 \times 10^{16}$ | $3.0 \times 10^{16}$ | -- | -- | $2.2 \times 10^{20}$ | $2.8 \times 10^{20}$ |
| $\Sigma_{dust}$[m$^{-2}$] | 25 | 9 | 16 | 9 | -- | -- | $2.8 \times 10^5$ | $8.8 \times 10^4$ |
| $Q_{dust}$[kg/s] | 0.6 | 0.5 | 2.7 | 3.2 | -- | -- | 1.5 | 1.0 |
| $\langle N_{CO2} \rangle$ [m$^{-2}$] | -- | -- | -- | -- | $1.75 \times 10^{10}$ | $1.32 \times 10^{10}$ | -- | -- |
| $\langle N_{CO} \rangle$[m$^{-2}$] | -- | -- | -- | -- | $1.70 \times 10^9$ | $1.26 \times 10^9$ | -- | -- |
| $Q_{CO2}$[s$^{-1}$] | -- | -- | -- | -- | $3.5 (+/- 0.9) \times 10^{24}$ | | -- | -- |
| $Q_{CO}$[s$^{-1}$] | -- | -- | -- | -- | $3.4 (+/- 2) \times 10^{23}$ | | -- | -- |

Using the analysis techniques as presented in Pittichova et al. (2008; see equations 4 and 5) we convert our 4.6 µm band fluxes to the number densities and production rates shown in Table 4, assuming a gas ejection velocity of 0.62 km/s. We scale the contribution from the two species for relative $CO_2$:CO abundance of roughly 10:1, similar to previous oppositions (Weaver et al. 1994, Bhardwaj & Raghuram 2011) to arrive at our relative production values. These represent the unique constraints on these species abundances for 103P at these distances. We have no available measurement of $H_2O$ production near the time of the WISE observations, so we are unable to provide reliable constraints on the $CO_2/H_2O$ ratio. However, extrapolation to May 10, 2010 from the production models as shown in Meech et al. (2011) and from rates observed on July 26, 2010 by Mumma et al. (2011), yields $H_2O$ values ~$3 \times 10^{26}$ mol/sec, suggesting that a derived $CO_2/H_2O$ ratio would be on the order of a few percent, i.e. a factor of a few lower than seen at perihelion during the EPOXI encounter (Weaver et al. 2011) and for previous orbits (Weaver et al. 1994). The $CO_2/H_2O$ ratio may possibly decrease as a function of heliocentric distance.

*Dust Trail:* The debris trail grain temperature ($T_g$) was calculated at each dMA from the ratio of the profile peak radiance in W3 to that in W4. The resulting temperatures were well represented over the range of trail distances ($2.26 < r < 2.39$ au) by the relation $T_g = 303 / r_h^{1/2} \pm 5$ K, where $r_h$ is given in AU, and is in good agreement with the relation derived for IRAS trail comets with multiband detections (Sykes, et al., 2005). As had been noted previously (Sykes, et al.,1990), trail temperatures are significantly (~10%) higher than the local equilibrium blackbody temperature. Possible explanations have been discussed extensively (Sykes et al.,1990, Sykes and Walker, 1992, Sykes et al.,2005); the least complex explanation assumes that randomly orientated trail grains maintain a latitudinal temperature gradient across their surfaces due to their very low thermal inertia. The optical depth, $\tau$, of the trail at the profile peak radiance was derived using the above relation for $T_g$ and assuming that the trail grains emit blackbody radiation, and is plotted in Figure 9 for each dMA observed. A solid line is drawn through the means of the W3 and W4 results, and represents the values of $\tau$ used to integrate the trail mass. The peak optical depths for 103P lie within the range of values (0.3 to 16) x $10^{-9}$ found for the 34 comets surveyed by Spitzer (Reach, et al., 2007) and (1.3 to 13.8) x $10^{-9}$ found for the 8 IRAS trails (Sykes and Walker, 1992).

The width of the trail, as derived from the FWHM of the radiance profile and the WISEcentric distance to the trail, is shown in Figure 9. The trail reaches a minimum width of ~65000 km slightly behind the comet at dMA = 0.12°, increases to a mean of ~140000 km ahead of the nucleus (-0.32° < dMA < -.16°), and increases to ~170000 km behind the nucleus at dMA = 1.04°. Our value of 65000 km near the nucleus is higher than the 40000 km width found from 22 µm Spitzer photometric imaging (Lisse, et al., 2009), however, this probably is due to the "contamination/averaging" previously discussed in the Analysis section. As such, it should be considered an upper limit to the trail width near the nucleus.

The velocity of trail grains can be deduced from measurements of the physical trail width (Sykes, et al., 1986). Applying their equations (1) and (2) to the above measurements of the width of the 103P trail and assuming isotropic perihelion emission, we derive a mean grain velocity relative to the nucleus of 3.6 ± 1.4 m/s. This is in good agreement with a radial velocity dispersion of the grains of 4 m/s from preliminary radar observations (Harmon et al., 2010).

The grain size is the final element required to calculate the mass of the observed debris trail. Whereas we cannot measure the grain size directly, we can measure the ratio $\beta$ = Force due to radiation/Force due to solar gravity = $1.14 \times 10^{-4} Q_{pr}/\rho d$ (Burns, et al., 1979) where $Q_{pr}$ = efficiency for radiation pressure ≈ 1.0 for grains with d » λ, ρ = grain density (g cm$^{-3}$), and d = grain diameter (cm). Trail grains ejected from the nucleus experience a reduced force of gravity due to solar radiation pressure that changes the semimajor axis of their orbits from that of the parent comet (Sykes and Walker, 1992). Figure 10 is a plot of the position of the peaks of the trail radiance profiles behind the nucleus vs dMA. The horizontal dashed line at pixel 100 is the position of the comet's orbit, that is, the zero velocity and zero β syndyne. Equation (2) from Sykes and Walker (1992) was used to calculate the positions of trail grains for a large set of β and find $\beta_{min}$ giving a minimum $\chi^2$ fit to the measured peak positions. The solid line is the zero velocity syndyne for $\beta_{min} = 1.0 \times 10^{-4}$. The dotted and dashed lines are for $\beta_{min} \pm 1\sigma$. In the case where $\rho = 1.0$ g/cm$^3$, $\beta_{min} = 1.0 \times 10^{-4}$ corresponds to a grain diameter = 1.1 cm, a diameter typical of dust trail particles (Sykes & Walker 1992, Reach et al. 2007), comet-associated meteoroids, and the icy grains imaged around the nucleus of 103P by the EPOXI spacecraft (A'Hearn et al. 2011). The $\pm 1\sigma$ $\beta_{min}$ values of $1.82 \times 10^{-4}$ and $2.10 \times 10^{-5}$ correspond to grain diameters of 0.63 and 5.4 cm respectively. Note that the density for large particles may be ~ 0.4 g/cm^3 (Richardson et al. 2007) and thus the diameter of the particles may be 2-3 times larger. This result agrees with preliminary radar Doppler spectra that show a broadband echo from > 1 cm diameter grains (Harmon, 2010).

Our solution for β also gives an estimate of the age of the trail particles. Grains observed at dMA=0.32 were emitted approximately 6.5 years (1 period) ago, while the grains observed most distant from the nucleus were emitted about 21 years (3.3 periods) ago.

103P is observed to have a significant trail of debris leading the comet. The minimum particle diameter ahead of the comet is that for which the effect of ejection velocity balances that of radiation pressure (Sykes, et al., 1986, Lisse et al. 2004). Setting the above referenced Equation (2) (Sykes and Walker, 1992) to zero and solving for β yields an estimate of the maximum value of $\beta = 1.91 \times 10^{-4}$ with corresponding minimum diameter of 0.59 cm in the leading portion of the trail (assuming $\rho = 1$).

The mass of the observed trail is estimated from the optical depth, peak normalized profile area, and $\beta$. If we assume that all the grains in a given field of view (FOV) are at the same temperature, then the optical depth, $\tau$, is just the fraction of the FOV area at the trail that is filled by the projected areas of the grains, thus $\tau = \pi n R^2 / (\alpha \Delta)^2$ where n is the number of grains of radius R within the FOV defined by the angular size of the pixel, $\alpha$, and the WISEcentric distance, $\Delta$. The mass of the grains within the FOV is just the mass of a single spherical grain times the number of grains, that is $M_{FOV} = 4/3 \, \pi R \rho \tau (\alpha \Delta)^2$ where $\rho$ is the grain density. In terms of $\beta$ and the WISE W4 pixel size of 5.5 arcsec this reduces to $M_{FOV} = 1.218 \times 10^{13} \, \tau \, \Delta^2 \, \rho/\beta$ (g/pixel$^2$) where $\Delta$ is in AU. The total mass of the observed trail is then $M_{total} = WLM_{FOV}$, where the width of the trail, W, in pixels is the area under the peak-normalized radiance profile, and L is the length of the trail, in pixels, observed on the sky. In practice, since $\tau$, $\Delta$, and W vary with sky position, $M_{total}$ was calculated as an integral along the trail length. Also, the portions of the trail ahead and behind the comet were treated separately, since each had a different value of $\beta$. This yields $3 \times 10^{12}$ grams for the total mass of the trail leading the nucleus, and $4 \times 10^{13}$ grams for the longer trail section in the following portion, assuming $\rho=1$ g/cm$^3$. As previously noted (Reach, et al., 2007), these values are lower limits to the trail mass because the full extent of the trail may not have been measured and larger grains may be present. The mass of the 103P trail is within the range of $3 \times 10^8$ to $3 \times 10^{11}$ kg found for the 8 trails observed by IRAS (Sykes and Walker, 1992) and the range $4 \times 10^7$ to $9 \times 10^{10}$ kg found for 6 Spitzer comets "with reasonable total mass estimates" (Reach, et al., 2007).

Given the age of the grains calculated above, we estimate the average mass loss rate of 103P meteoroids over 21 years is 62 kg/s into the trail, which is greater by a factor of 9 than the average orbital mass loss rate of ~7 kg/s into coma dust. This rate would amount to the loss per orbit of ~ %1.5 of the mass of the comet's nucleus, assuming a density near 1 g/cm$^3$. Our value falls within the range of 4 to 250 kg/s found for the IRAS trails, but outside the range of 0.2 to 36 kg/s for the Spitzer trails. If we exclude 2P/Encke and 29P/Schwassmann-Wachmann 1 from the IRAS list, then the IRAS range of mass loss rates for more typical Jupiter family comet trails is 4 to 20 kg/s, totally within the range measured by Spitzer, and our value of 62 kg/s for 103P is now well outside that range.

## Conclusions

Observed by WISE at a unique time and heliocentric distance relative to its spacecraft encounter, 103P exhibits interesting characteristics for a short period comet. These data help constrain behavior over a longer period than the immediate pre and post-encounter observations, and the simultaneous and near-simultaneous data presented here in combination yield the following results:

- The extracted thermal flux from the nucleus indicates a mean diameter consistent with encounter measurements for the geometric mean radius of the body. This is as expected, considering the ~38-hour span over which the 18 WISE images were taken covers two full rotations of the 18-hour period of the comet. The fitted η values of ~1.2 are marginally higher than seen in other data sets.

- The $CO_2$ (and CO) detected in W2 excess is consistent with $CO_2$ production rates on the order of $3.5 \times 10^{24}$ molecules per second on 10 May 2010 at $r_h$ = 2.3 AU. This is lower than the estimated production seen near perihelion, at $r_h$ = 1.06 AU (Meech et al. 2011, Weaver et al. 2011), but still plays a significant role in the cometary activity at these distances.

- A Black Body curve fit to the observed thermal flux from coma dust particles yields dust temperatures near the temperature for a black body at the comet's 2010 May 10 heliocentric distance, 2.3 AU, and is consistent with large-particle (> 11 μm) dust dominance.

- The extended trail's extracted peak radiance locations were fit with an array of dust β parameter values. The best-fit values are consistent with large-particle sizes on the order of centimeters, and with a long-lived dust trail that is extant for periods greater than a single orbit.

- The total trail mass and mass production rate coupled with a small nuclear diameter indicate that 103P's refractory component dominates its mass.

The WISE 103P data provides meaningful values for each observable that the data provided, and serves as a good example of what the WISE cometary body data set is capable of yielding on individual bodies, and for statistical comparisons of the over 120 comets observed in the sample.

## Acknowledgements

This publication makes use of data products from the Wide-field Infrared Survey Explore, which is a joint project of the University of California, Los Angeles, and the Jet Propulsion Laboratory/California Institute of Technology, funded by the National Aeronautics and Space Administration. This publication also makes use of data products from NEOWISE, which is a project of the Jet Propulsion Laboratory/California Institute of Technology, funded by the Planetary Science Division of the National Aeronautics and Space Administration. Observing time was allocated at the University of Hawaii 88-inch telescope at Mauna Kea Observatory, the Palomar Observatory Hale 200-inch telescope, by the National Optical Astronomy Observatory at the SOAR telescope, and on Steward Observatory's 2.3m on Kitt Peak. The SOAR Telescope is a joint project of: Conselho Nacional de Pesquisas Cientificas e Tecnologicas CNPq-Brazil, The University of North Carolina at Chapel Hill, Michigan State University, and the National Optical Astronomy Observatory. The Hale Telescope at Palomar Observatory is operated as part of a collaborative agreement between the California Institute of Technology, its divisions Caltech Optical Observatories and the Jet Propulsion Laboratory (operated for NASA), and Cornell University. J. Bauer would also like to thank Dr. M. Hanner for her valuable advice on the analysis, and the anonymous reviewer for their helpful comments.

**Figure Captions**

**Figure 1** A three-color composite image of 103P/ Hartley 2 from the WISE data taken May 10-11, 2010. The 18 stacked 4.6, 11.6, and 22.1 μm images were mapped to blue, red, and green channels 22 arcmin on a side, insets of the individual band images close to the comet, 5 arcmin on a side, are shown below (W2, W3, and W4 images, left to right). The sky-projected anti-solar vector is indicated by the red dashed arrow, and the projected anti-velocity vector by the orange dotted arrow.

**Figure 2** The optical image taken at the UH 2.2m on the night of May 23$^{rd}$, 2010. The image was taken with the Tek 2048 camera, .22"/pixel, and shown here with 3 arcmin on a side. The sky-projected anti-solar vector is indicated by the dashed arrow, and the projected anti-velocity vector by the dotted arrow.

**Figure 3** Coma Temperature fit to the 22 arcsecond aperture thermal photometry in the two longest WISE wavelength bands. Nucleus and reflected light signal were subtracted from the flux in the two bands prior to fitting. The 4.6 μm band coma signal is also shown, along with the R-band brightness measured on May 23$^{rd}$, adjusted for distance, and with a small 4% adjustment for the projected activity based on multiple observations (see text). Two reflected-light models are shown, one with a neutral reflectance (heavy dotted) and one with a reddening law based on Jewitt & Meech (1986) averaged out to 3.5 μm (light dotted line), i.e. with flux proportional to the wavelength to the approximately 0.2 power. The uncertainties to the temperature fit are on the order of +/- 9K, and the fitted temperature (195K) closely matches the black body temperature for that distance

(185K). Excess W2 flux cannot be explained by thermal and reflected-light contributions of dust.

**Figure 4** Thermal models of extracted nucleus signal. Figure 4A shows the W3 (left) and W4 (right) residual signal after coma extraction, using 1/r model coma fits. Figure 4B shows the thermal fits to the data. Unconstrained fits for beaming parameter ($\eta$) values yield fits of $\eta\sim2.5$ (solid line), and diameter values of 1.7km. $\eta$ parameter values constrained to 2.0 (dot-dashed line), 1.2 (dashed line) and 0.94 (dotted line) yield diameters of 1.5, 1.1 and 0.9km, respectively.

**Figure 5** Particle Size Distribution (PSD). Log number is shown vertically, while log mass is shown on the bottom scale and the corresponding grain radius size, in microns, is shown on the scale above. The 103P data derived number of particles in the 11 arcsecond aperture radius (pentagon) and 22 arcsecond aperture (cross) are shown. The flux contributions from the nucleus have been subtracted. For comparison, Deep Impact particle densities (triangles), and Echeclus particle numbers (diamonds and squares, as presented in Bauer et al. 2008) are also shown. Stardust PSD slope ($\alpha = -0.75$, in log N/log kg units, where N is the estimated total number of dust grains in the aperture) is shown as the dashed line, rescaled from dust fluence values to an aperture encompassing a similar $\rho$ size. Echeclus' PSD best-fit ($\alpha = -0.87$) is shown as a dotted line, and the solid line is the best fit to 103P PSD data ($\alpha = -0.97$).

**Figure 6A** W3 (top) and W4 (bottom) images shifted, rotated, and stacked with respect to delta mean anomaly and centered on dMA = $0.2^\circ$. The W3 image is a stack of 17 frames, while the W4 image is a stack of 15 frames. The trail is seen in both images as a faint diffuse horizontal line through the center of the images. Note the background stars running diagonally through the trail image.

**Figure 6B** Radiance profiles for the stacked W3 (left) and W4 (right) images shown in Figure 6A. Plotted points are the mean radiances for each row. The units are data numbers (DN) given in the WISE level 1b images. The solid line is the best-fit Moffat function.

**Figure 7** W3 and W4 peak radiance (MJy/sr) versus dMA (-dMA is ahead of the nucleus). The dotted lines bound the region of dMA where the trail radiance is contaminated by coma and tail emission. The maximum contamination is ~10% of the peak. Error bars are $\pm 1\sigma$.

**Figure 8** The optical depth at the peak of the radiance profile is shown. The solid line is drawn through the means of the W3 and W4 values, and defines the values used for the integration of trail mass. Error bars are $\pm 1\sigma$.

103P, Bauer et al. 2011

**Figure 9** The width of the trail versus dMA. The widths plotted are derived from the FWHM of the radiance profiles and are corrected for the distance from WISE to the trail. Error bars are ±1σ.

**Figure 10** The positions of the peaks (+ symbols) of the radiance profile as a function of dMA. Error bars are ±1σ. The horizontal dashed line at pixel 100 is the position of the comet's orbit, that is, the zero velocity and zero β syndyne. The solid line is the zero velocity syndyne for $\beta_{min}$ = 1.01x10$^{-4}$. The dotted and dark-dashed lines are the $\beta_{min}$ ± 1σ values.

# Figures

103P, Bauer et al. 2011

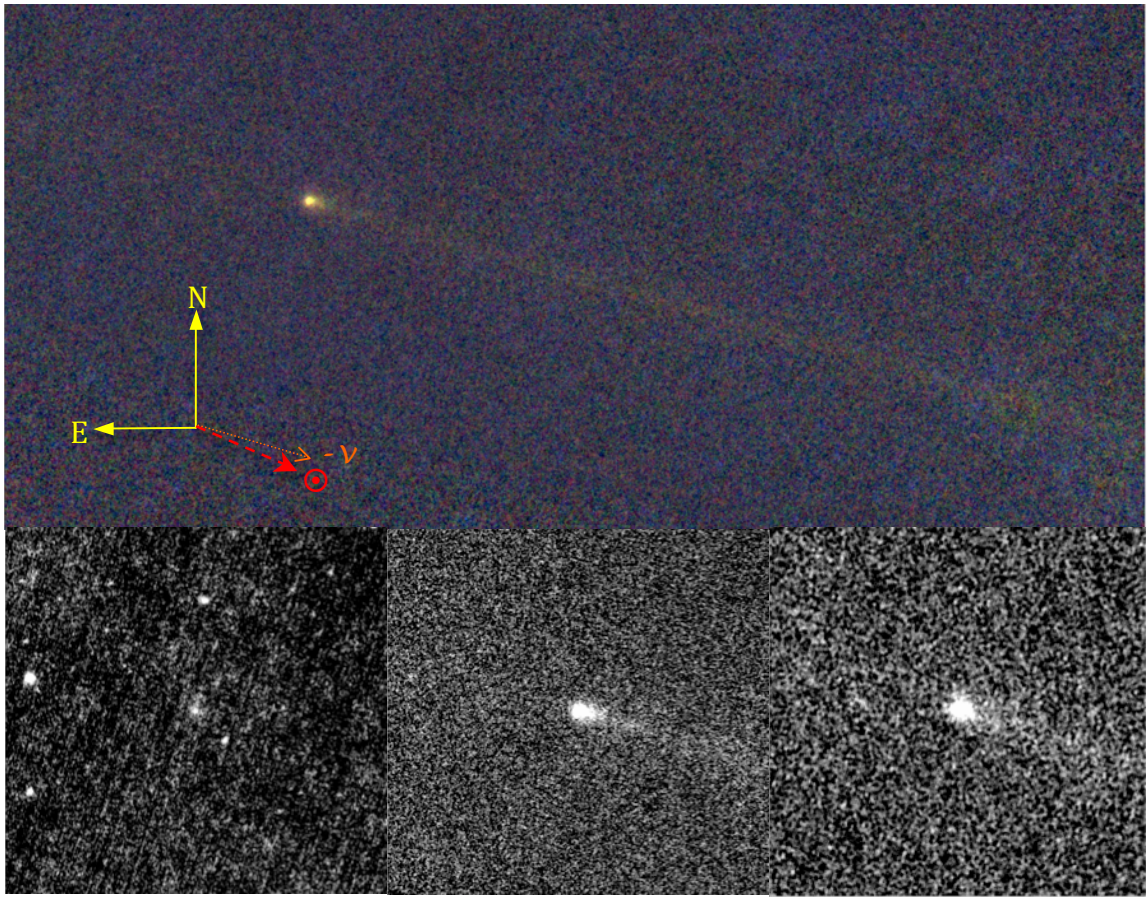

Figure 1

103P, Bauer et al. 2011

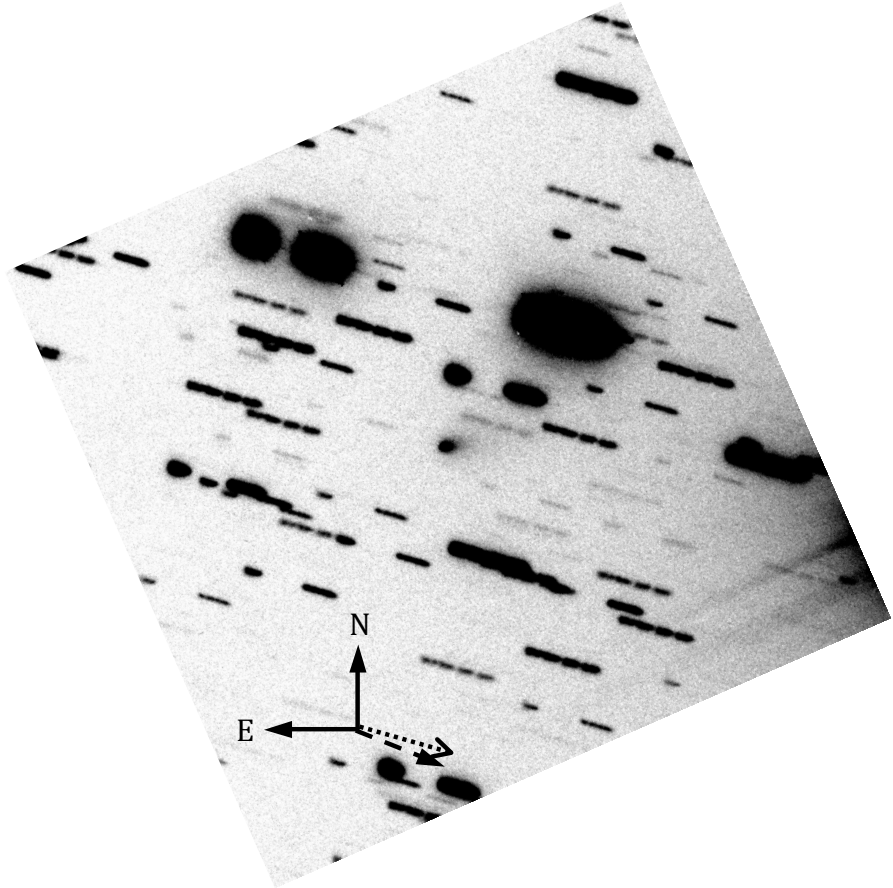

Figure 2

103P, Bauer et al. 2011

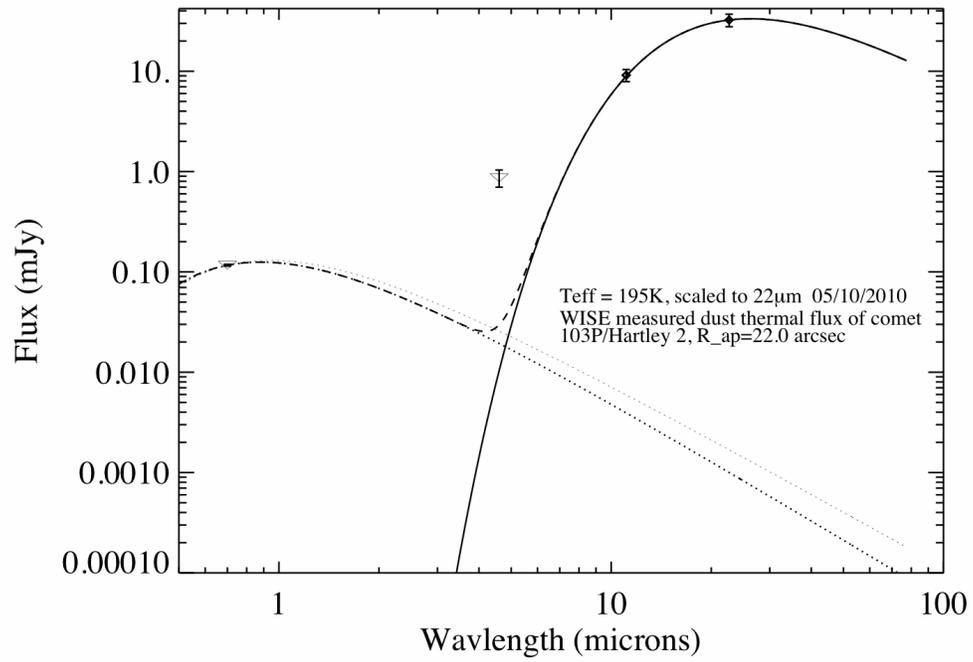

Figure 3

103P, Bauer et al. 2011

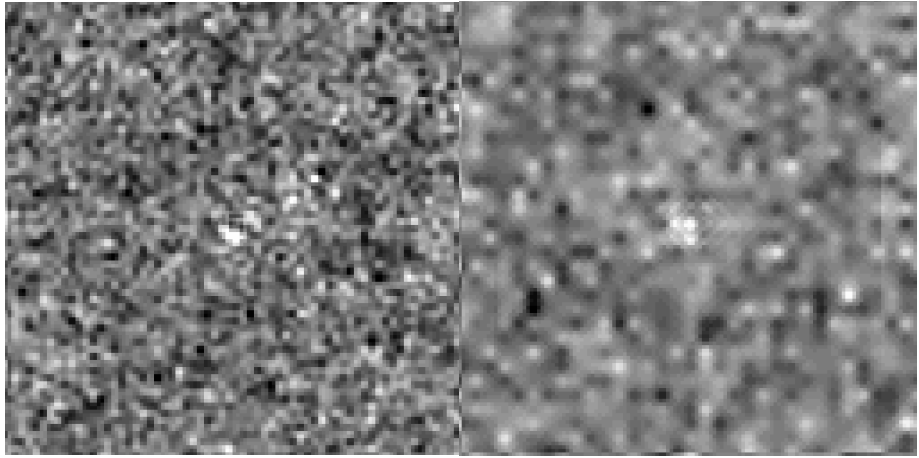

Figure 4A

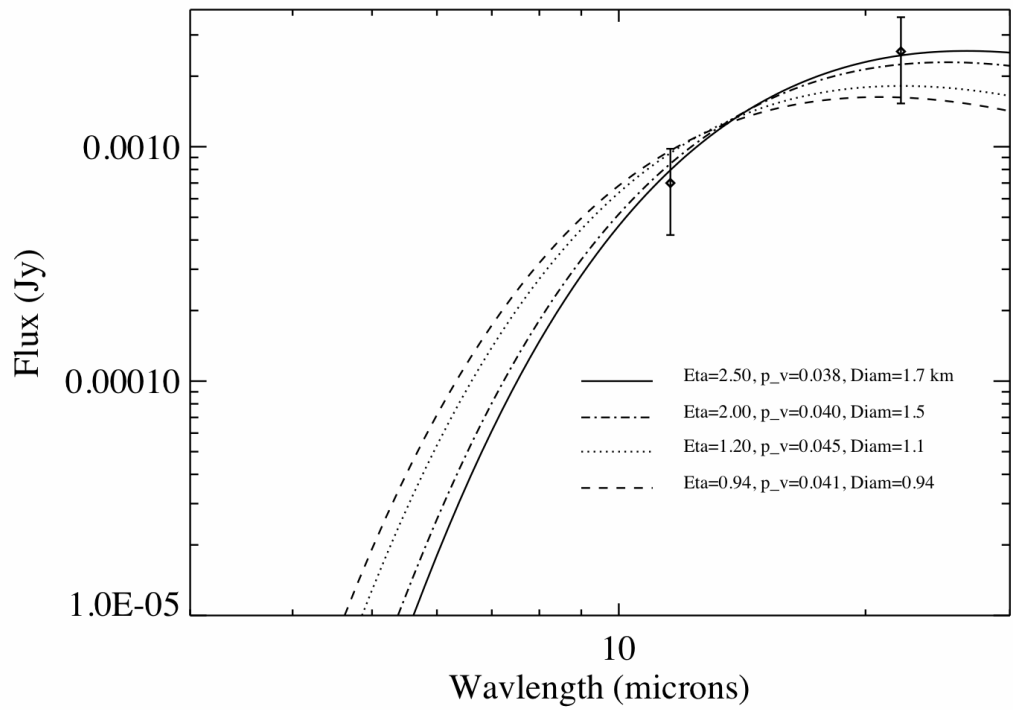

Figure 4B

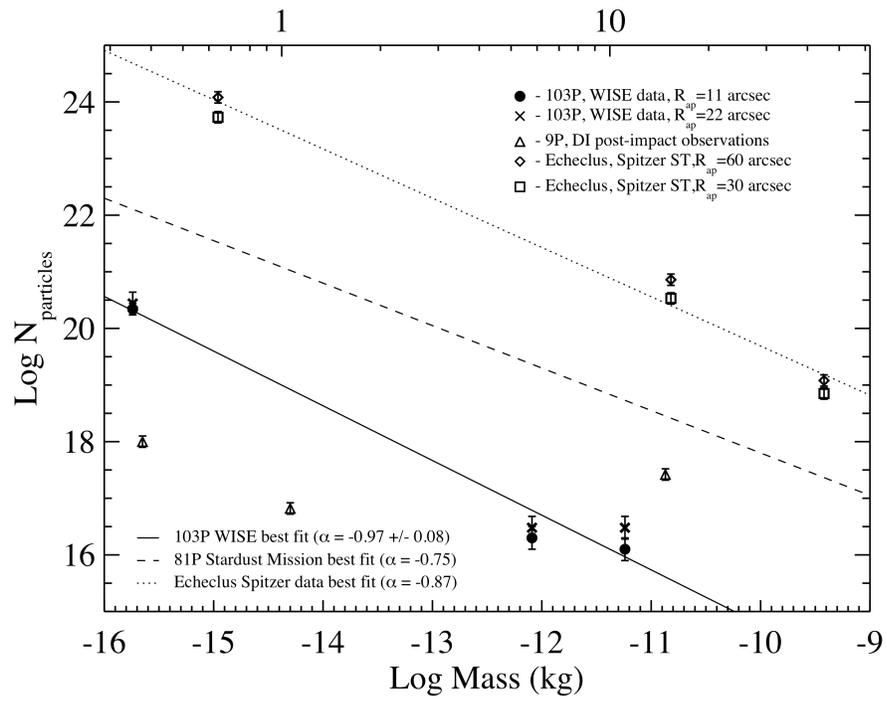

Figure 5

103P, Bauer et al. 2011

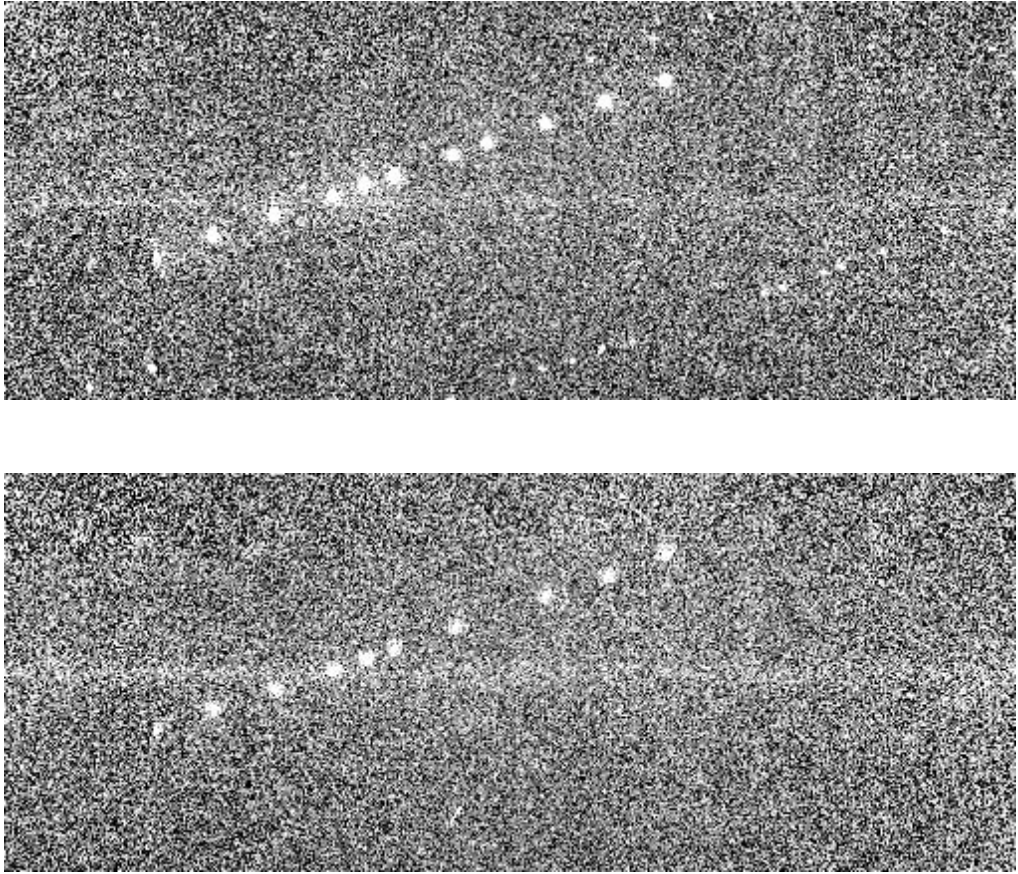

Figure 6A

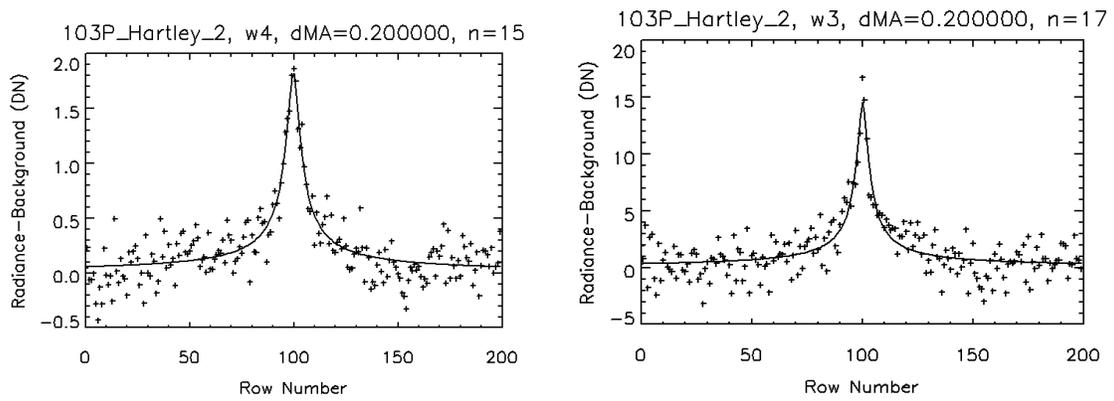

Figure 6B

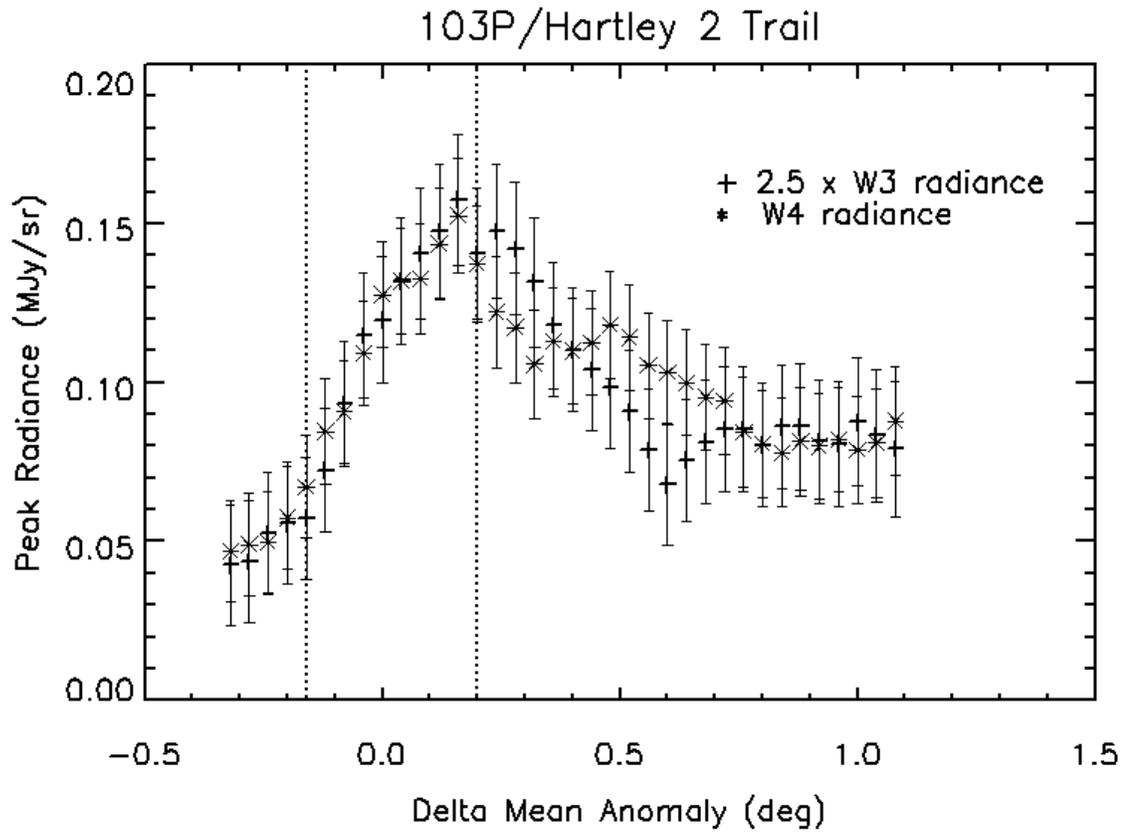

Figure 7

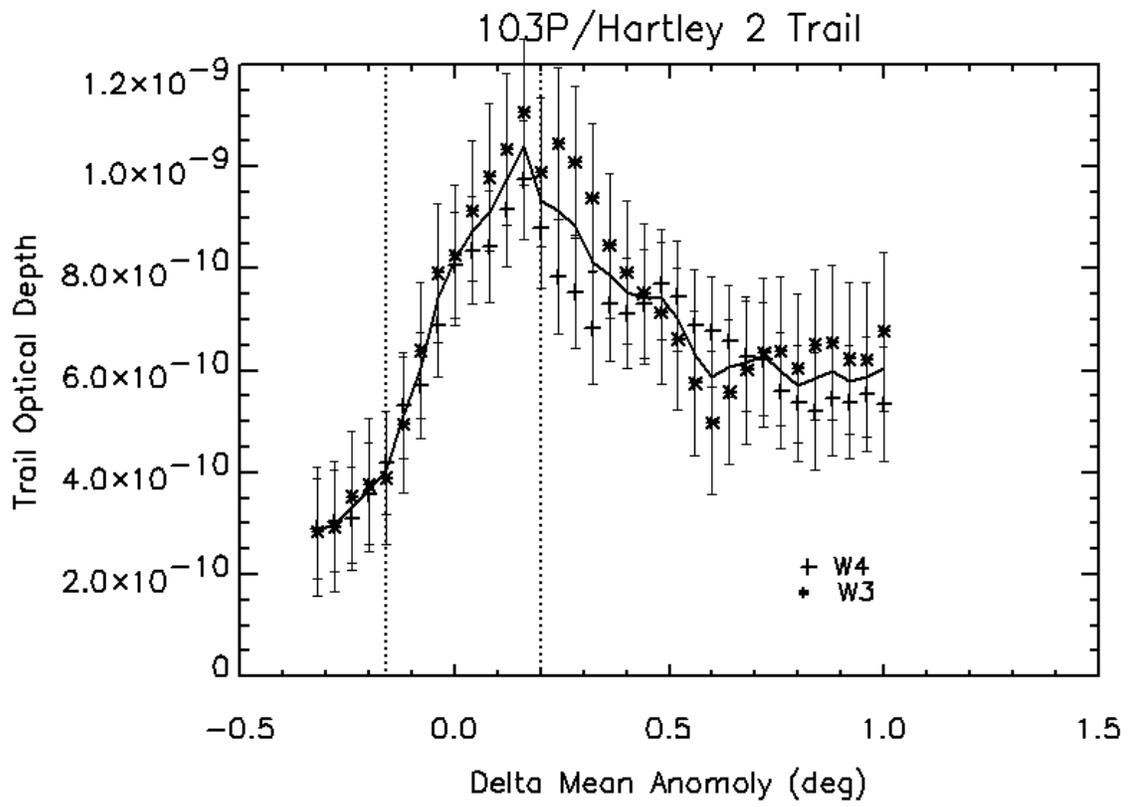

Figure 8

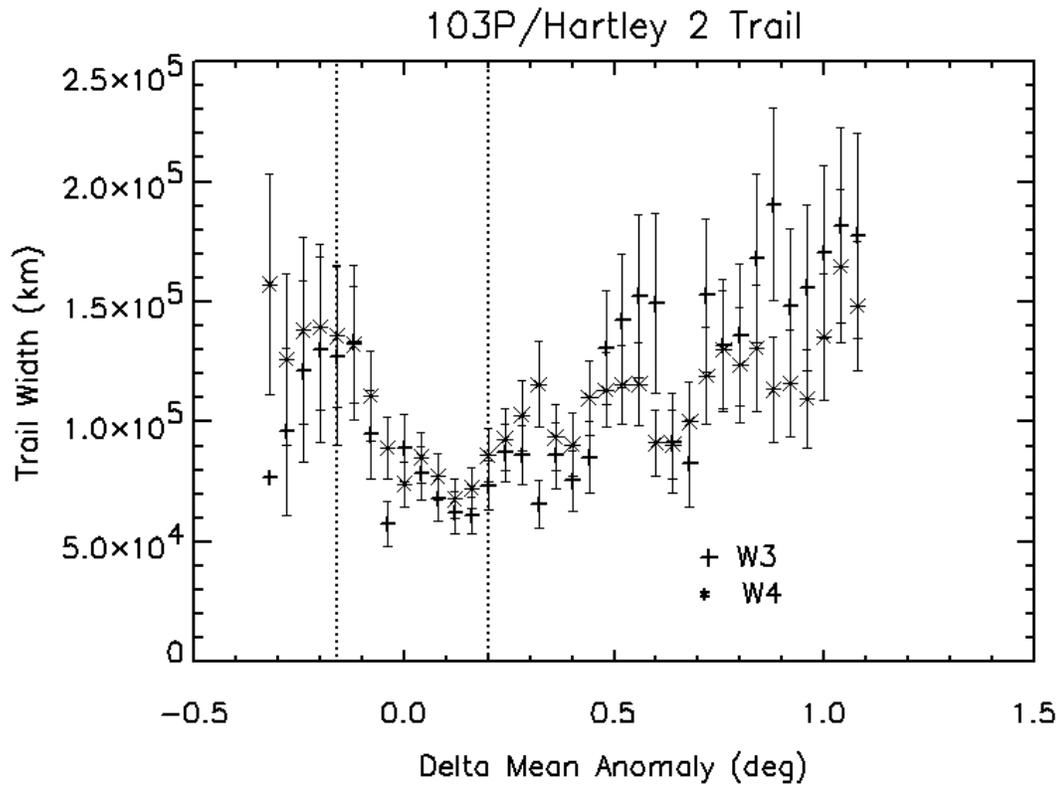

Figure 9

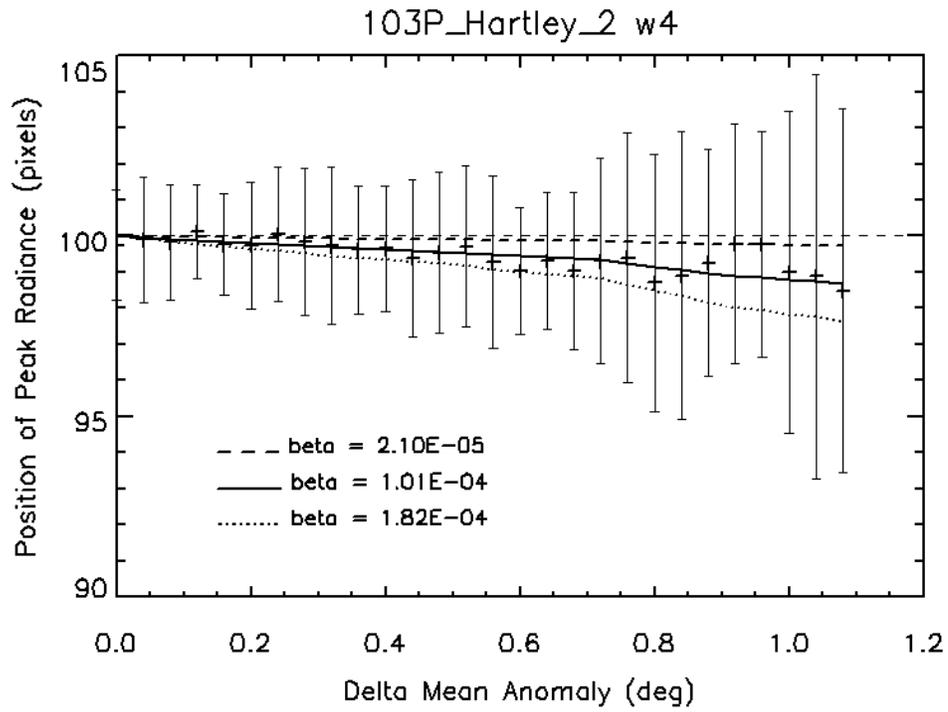

Figure 10